\documentclass[12pt]{article}
\usepackage[dvips]{epsfig}

\begin{document}

\begin{center}
{\bf The Color Dipole Picture and the Ratio of the Longitudinal 
to the Transverse Photoabsorption Cross Section \footnote{
Presented at Diffraction 2008, La Londe-les-Maures, France, September
9 - 14, 2008} 
}
\end{center}


\begin{center}
{\bf Dieter Schildknecht} \footnote{email: 
Dieter.Schildknecht@physik.uni-bieleld.de}

Fakult\"at f\"ur Physik, Universit\"at
      Bielefeld, D-33615 Bielefeld, Germany     \\
    and \\
    Max-Planck Institut f\"ur Physik (Werner-Heisenberg-Institut)
F\"ohringer Ring 6, D-80805 M\"unchen, Germany 
\end{center}

\begin{abstract}
 The transverse size of $q \bar q$ fluctuations of a longitudinally polarized
photon is reduced relative to the transverse size of $q \bar q$ fluctuations
of a transversely polarized photon. This implies a model-independent 
prediction of the ratio $R(W^2 , Q^2) \equiv \sigma_L / \sigma_T = 0.375$, 
or, equivalently, $F_L / F_2 = 0.27$, for $x \cong Q^2 / W^2 \ll 1$ and $Q^2$
sufficiently large, while $R(W^2 , Q^2) = 0.50$, if this effect is ignored. 
Experimental data from HERA confirm the transverse-size reduction.
\end{abstract}


At low values of $x \cong Q^2 / W^2 \ll 1$, in terms of the imaginary part of the
virtual Compton-forward-scattering amplitude, deep inelastic 
scattering proceeds via diffractive forward scattering of timelike hadronic 
\cite{1,2} quark-anti-quark, $q \bar q$, fluctuations of the virtual spacelike
photon on the proton. In its interaction with the proton, a $q \bar q$ 
fluctuation acts as a color dipole \cite{3}. 
A massice $q \bar q$ fluctuation is identical to the $(q \bar q)^{J=1}$ vector 
state originating from a timelike photon in $e^+ e^-$ annihilation at an 
$e^+ e^-$ energy equal to the mass, $M_{q \bar q}$, of the $q \bar q$ state. 

The coupling strength of a timelike photon of mass $M_{q \bar q}$ to a 
$q \bar q$ state of mass $M_{q \bar q}$ is determined by the 
longitudinal and transverse components of the electromagnetic current 
\cite{4,5},
\begin{equation}
\sum_{\lambda = - \lambda^\prime = \pm 1} \vert j_L^{\lambda, \lambda^\prime}
\vert^2 = 8 M^2_{q \bar q} z (1-z) ,
\label{1}
\end{equation}
and
\begin{equation}
\sum_{\lambda = - \lambda^\prime = \pm 1} \vert j_T^{\lambda, \lambda^\prime}
(+) \vert^2 = \sum_{\lambda = - \lambda^\prime = \pm 1} 
j_T^{\lambda, \lambda^\prime} (-) \vert^2 = 2 M^2_{q \bar q} (1 - 2 z (1-z)),
\label{2}
\end{equation}
where $j_T^{\lambda , \lambda^\prime} (+)$ and 
$j_T^{\lambda , \lambda^\prime} (-)$ stand for the positive and negative
helicity of the transversely polarized photon. Longitudinal and 
transverse in (\ref{1}) and (\ref{2}) refer to the $\gamma^* p$ axis. We
imagine a situation in which a timelike photon, upon dissociating into a 
$q \bar q$ pair, interacts with the proton, and we define the $\gamma^* p$ axis
by the $\gamma^*$ and proton three-momenta in e.g. the $\gamma^* p$  
center-of-mass frame. In (\ref{1}) and (\ref{2}), the variable $0\le z \le 1$
denotes the usually employed momentum-fraction variable that is related to the 
$q \bar q$-rest-frame angle between the $\gamma^* p$ axis and the 
three-momentum of the (massless) quark by 
\begin{equation}
    \sin^2 \vartheta = 4 z (1-z).
\label{3}
\end{equation}
The $q \bar q$ mass, $M_{q \bar q}$, in terms of the transverse momentum 
squared, $\vec k_\bot^2$, of the quark (antiquark) and $z (1-z)$ is 
given by 
\begin{equation}
M^2_{q \bar q} = \frac{\vec k^2_\bot}{z (1-z)}  .
\label{4}
\end{equation}
According to (\ref{1}), (\ref{2}) and (\ref{4}), longitudinal photons of fixed
mass, $M_{q \bar q}$, produce dominantly $q \bar q$ pairs of relatively large 
$| \vec k_\bot |$, while transverse photons lead to $q \bar q$ pairs of
dominantly small $|\vec k_\bot|$. Quantitatively, one finds that the ratio
of the average transverse momenta is given by \cite{5}
\begin{equation}
\rho = \frac{\langle \vec k_\bot^2 \rangle_L}{\langle \vec k_\bot^2 
\rangle_T} = \frac{4}{3}. 
\label{5}
\end{equation}
From the uncertainty relation, the ratio of the effective transverse sizes of 
the $(q \bar q)^{J=1}_{L,T}$ states becomes \cite{5}
\begin{equation}
\frac{\langle \vec r_\bot^2 \rangle_L}{\langle \vec r_\bot^2 \rangle_T} =
\frac{1}{\rho} = \frac{3}{4}.
\label{6}
\end{equation}
Longitudinal photons produce ``small-size'' pairs, 
$(q \bar q)^{J=1}_L$, while transverse photons produce ``large-size'' pairs, 
$(q \bar q)^{J=1}_T$. The ratio of the sizes is given by (\ref{6}). 

The transition from a timelike photon interacting with the proton via a 
$q \bar q$ pair of fixed mass $M_{q \bar q}$ to a spacelike photon fluctuating
into a continuum of massive $q \bar q$ (vector) states is provided by the
color-dipole picture (CDP) \cite{3}. 
In a representation of the CDP that explicitly expresses the total 
photoabsorption cross section in terms of the scattering of $(q \bar q)^
{J=1}_{L,T}$ longitudinal and transverse $(J=1)$ vector states, one has 
\cite{6,5}
\begin{equation}
\sigma_{\gamma^*_{L,T}} (W^2, Q^2) = \frac{2\alpha R_{e^+ e^-}}{3\pi^2} Q^2 \int
d^2 r^\prime_\bot K^2_{0,1} (r^\prime_\bot Q) \sigma_{(q \bar q)^{J=1}_{L,T}p}
(r^\prime_\bot , W^2),
\label{7}
\end{equation}
where $Q \equiv \sqrt{Q^2}, R_{e^+ e^-} \equiv 3 \sum_q Q^2_q$, and $r^\prime_
\bot$ is related to the transverse quark- antiquark separation by
\begin{equation}
\vec r^{~\prime}_\bot = \vec r_\bot \sqrt{z(1-z)} .
\label{8}
\end{equation}
The representation (\ref{7}) is called the ``$r^\prime_\bot$-representation''. 
It is based on an explicit factorization of the $z(1-z)$ dependence of the
$\gamma^* q \bar q$ couplings of longitudinal and transverse photons in
(\ref{1}) and (\ref{2}). The integration over the contributing mass 
continuum in (\ref{7}) appears as integration over $\vec r^{~\prime}_\bot$. 
The transverse-size enhancement (\ref{6}) enters (\ref{7}) via
\begin{equation}
\sigma_{(q \bar q)^{J=1}_T p} (r^\prime_\bot , W^2) = \rho 
\sigma_{(q \bar q)^{J=1}_L p} (r^\prime_\bot , W^2).
\label{9}
\end{equation}

Due to the strong decrease of the modified Bessel functions $K_{0,1} (r^\prime_
\bot Q)$ in (\ref{7}) with increasing argument, $r^\prime_\bot Q$, for 
sufficiently large $Q^2$, the integrals in (\ref{7}) are dominated by the 
$r^{\prime 2}_\bot \rightarrow 0$ behavior of the dipole cross sections. The
coupling of the $q \bar q$ pair to two gluons implies vanishing of the 
interaction proportional to $r^{\prime 2}_\bot$, 
\begin{equation}
\sigma_{(q \bar q)^{J=1}_{L,T}} (r^\prime_\bot , W^2) \sim r^{\prime 2}_\bot , 
\,\,\,\,\,\,\,\,\,\,(r^{\prime 2}_\bot \rightarrow 0).
\label{10}
\end{equation}
From (\ref{7}) and (\ref{9}) together with ``color transparency'' (\ref{10}), 
for sufficiently large $Q^2$ and $x \ll 1$, we have \cite{5}
\begin{equation}
R (W^2 , Q^2) \equiv \frac{\sigma_{\gamma^*_L p} (W^2 , Q^2)}{\sigma_{
\gamma^*_T p} (W^2 , Q^2)} = \frac{\int d^2 r^\prime_\bot 
r^{\prime 2}_\bot K^2_0 (r^\prime_\bot Q)}{\rho \int d^2 r^\prime_\bot
r^{\prime
2}_\bot K^2_1 (r^\prime_\bot Q)} = \frac{1}{2\rho}  .
\label{11}
\end{equation}
A measurement of $R(W^2, Q^2)$ at low $x$ and sufficiently large $Q^2$ 
provides a determination of the ratio $1 / \rho$ of the cross sections of 
longitudinal and transverse $q \bar q$ fluctuations on the proton, compare 
$\rho$ in (\ref{9}). 

In fact, the reduced cross section of DIS, for sufficiently large $Q^2$ and 
$x \ll 1$, employing (\ref{11}), may be directly expressed in terms of $\rho$,
\begin{equation}
\sigma_r (x, y, Q^2) = F_2 (x , Q^2) \left( 1 - 
\frac{y^2}{1+(1-y)^2} \frac{1}{1+2\rho} \right) .
\label{12}
\end{equation}
In (\ref{12}), $y=Q^2 / xs$. A variation of the $ep$ center-of-mass energy
$\sqrt s$ at fixed $Q^2$ and $x$ yields $1/(1+2\rho)$ from the slope of a 
straight-line fit of 
$\sigma_r (x,y,Q^2)$ against $y^2 / (1+(1-y)^2) \le 1$. 

With $\rho = 4/3$ from the transverse-size enhancement of transversely 
polarized $(q \bar q)^{J=1}_T$ states relative to longitudinally polarized
$(q \bar q)^{J=1}_L$ states, $R(W^2, Q^2)$ from (\ref{11}) becomes 
\begin{equation}
R (W^2 , Q^2) = \frac{3}{8} = 0.375, \,\,\,\,\,\,\,\,\,\, (\rho = \frac{4}{3}) .
\label{13}
\end{equation}
It is of interest to compare this result (\ref{13}) with the result 
from an assumed helicity independence, $\rho = 1$, based on an equality of 
scattering cross sections on the proton of $(q \bar q)^{J=1}_h$ 
fluctuations for helicities $h=0, h=1$ and $h=-1$. In the case of helicity 
independence, we have 
\begin{equation}
R(W^2, Q^2) = 0.5, \,\,\,\,\,\,\,\,\,\, (\rho = 1)
\label{14}
\end{equation}
in distinction from (\ref{13}). 

In terms of the proton structure functions, $F_2 (W^2 , Q^2)$ and 
$F_L (W^2 , Q^2)$, the result (\ref{11}) becomes
\begin{equation}
\frac{F_L (W^2 , Q^2)}{F_2 (W^2 , Q^2)} = \frac{1}{1 + 2 \rho} = 
\left\{ \matrix{ \frac{3}{11} \cong 0.27, & {\rm for} \,\, \rho = 
\frac{4}{3} , \cr
\frac{1}{3} \cong 0.33 , & {\rm for} \,\, \rho = 1 } \right.  .
\label{15}
\end{equation} 

In fig. 1, we show a comparison \cite{7,8} of the prediction (\ref{15}) for 
$F_L (W^2, Q^2)$ in terms of $F_2 (W^2, Q^2)$ for the transverse size
enhancement of $\rho = 4/3$ with experimental data for $F_L (W^2, Q^2)$. 
For $F_2 (W^2, Q^2)$, the results from the H1 PDF2000 fit were used as 
an imput-parameterization of the experimental data for $F_2 (W^2, Q^2)$. 
We predict a slightly larger contribution to $F_2 (W^2, Q^2)$ of 
longitudinal photons than obtained from the H1 PDF2000 fit. Helicity
independence, $\rho = 1$, is disfavored, since it shifts our prediction for 
$F_L (W^2, Q^2)$ upwards by about 22\%. A plot of the experimental values
of $\rho$ directly deduced from the measured slopes of the reduced cross 
section would be rewarding, since it would provide a direct test of the 
transverse-size enhancement of transversely polarized relative to 
longitudinally polarized $q \bar q$ fluctuations. 

In fig. 2 \cite{9}, we show the result of an analysis of H1 data that relied 
on QCD predictions to extract the longitudinal photoabsorption 
cross section from
the measured reduced cross section. The consistency of the H1 PDF2000 fit 
with the separation data in fig. 1 a posteriori justifies this method. The 
theoretical predictions in fig. 2 were based on our explicit ansatz \cite{10}
for the dipole cross section that was based on $\rho = 1$. The inclusion of the 
transverse size enhancement, $\rho = 4/3$, implies a downward shift of the 
theoretical curves by a factor of approximately 0.82 that improves 
agreement with the data. 

\begin{figure}
\epsfysize=8cm
\centering{\epsffile{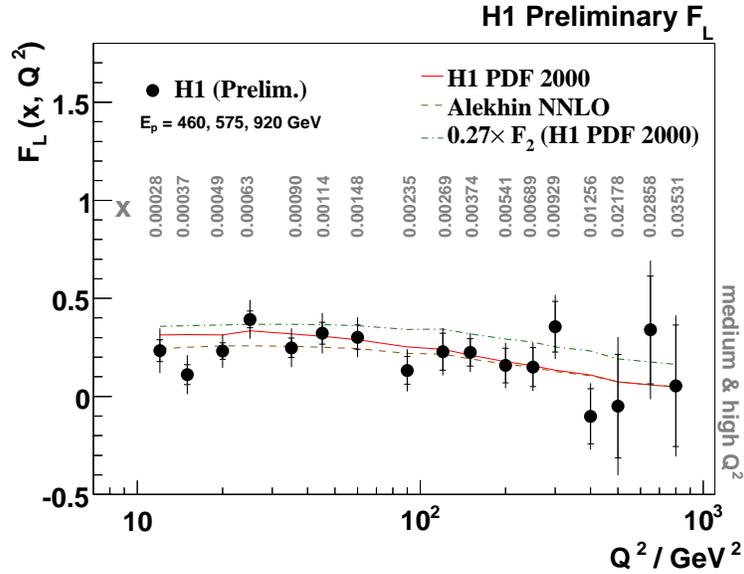}} 
 \caption{Experimental data from HERA for the longitudinal proton 
structure function as a function of $Q^2$ for various values of 
$x\ll 1$ compared with theoretical predictions \cite{7}.}
\end{figure}

\begin{figure}
\epsfysize=8cm
\centering{\epsffile{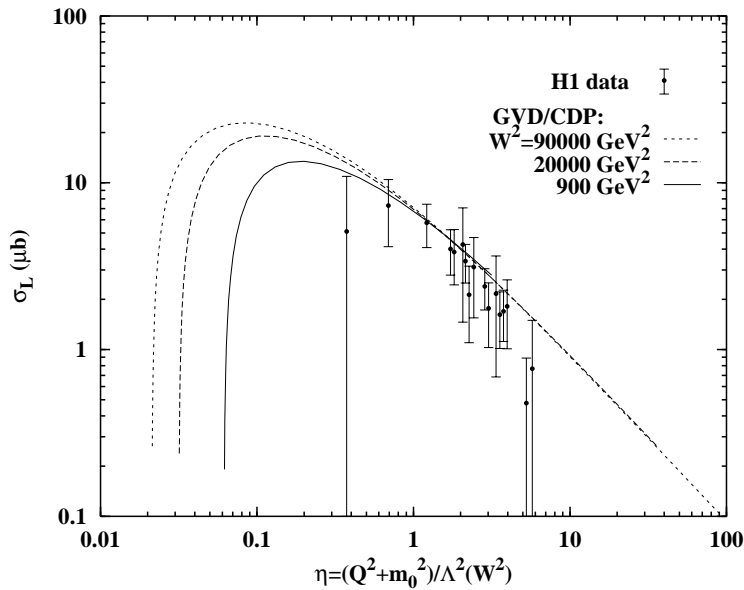}} 
 \caption{The longitudinal photoabsorption cross section as a function
of the scaling variable $\eta$ compared with theoretical predictions \cite{9}.}
\end{figure}

An interesting upper bound on $R(W^2, Q^2) \le 0.37248$ was recently 
found \cite{11} in the CDP under the frequently employed assumption 
that the dipole 
cross section is independent of $z(1-z)$. The fact that the bound is close 
to our prediction (\ref{13}) is a numerical accident. Experimental values
below the bound neither require nor rule out a dependence on $z(1-z)$. Such 
a dependence is present, if the kinematic domain of validity of the CDP is 
extended by imposing an upper bound \cite{6} on the masses (\ref{4}) of 
the contributing $q \bar q$ fluctuations. 

In conclusion, we stress that our prediction of $R(W^2, Q^2)$, or of the 
ratio $F_L / F_2$ of the structure functions, is independent of a 
specific model 
for the dipole cross section. The prediction rests on the CDP combined with the 
transverse-size enhancement (of transversely polarized $(q \bar q)^{J=1}$ 
fluctuations relative to longitudinally polarized ones) and color 
transparency. The HERA experimental data confirm this prediction and disfavor
the hypothesis that $(q \bar q)^{J=1}$ fluctuations interact independently
of their helicity. 

\bigskip
\noindent
{\bf Acknowledgement}\\

Many thanks to Kuroda-san for a long-standing collaboration and to 
Allen Caldwell and Vladimir Chekelian for useful discussions on the 
HERA data. Last not least, many thanks to the organizers for a 
fruitful workshop Diffraction 2008. \\
This work was supported by Deutsche Forschungsgemeinschaft, contract
number schi 189/6-2.


\bibliography{sample}




\end{document}